\def\be{\begin{equation}}
\def\te{\end{equation}}
\def\bea{\begin{eqnarray}}
\def\nn{\nonumber}
\def\tea{\end{eqnarray}}

\def\ha{{1\over 2}}

\def\xib{\overline{\xi}}

\def\a{\alpha}
\def\b{\beta}

\def\k{\kappa}

\def\m{\mu}

\def\n{\nu}

\def\O{\Omega}

\def\bb{\bibitem}
\def\mb{\overline{m}}

\input{epsf}
\documentstyle[prl,aps,twocolumn]{revtex}

\catcode`\@=11

\def\maketitle2{\par 
\begingroup
\let\cite\@bylinecite
\def\thefootnote{\fnsymbol{footnote}}%
\twocolumn[\@maketitle2\vskip2pc]%
\thispagestyle{plain}\@thanks
\endgroup
\def\thefootnote{\arabic{footnote}}%
\setcounter{footnote}{0}%
\let\maketitle2\relax \let\@maketitle2\relax
\let\@thanks\relax \let\@authoraddress\relax \let\@title\relax
\let\@date\relax \let\thanks\relax \let\@abstract\relax 
\let\@pacs\relax}

\def\abstract#1{\gdef\@abstract{{\par 
\bgroup
\ifdim\prevdepth=-1000pt \prevdepth0pt\fi
\hsize\columnwidth
\dimen0=-\prevdepth \advance\dimen0 by17.5pt \nointerlineskip
\small\vrule width 0pt height\dimen0 \relax}{~~}#1\egroup}}

\def\pacs#1{\gdef\@pacs{{\par 
\bgroup
\hsize\columnwidth \parindent0pt
\ifdim\prevdepth=-1000pt \prevdepth0pt\fi
\dimen0=-\prevdepth \advance\dimen0 by20pt\nointerlineskip
\egroup} PACS numbers:~#1}}

\def\@maketitle2{
\@preprint
\@title
\ifdim\prevdepth=-1000pt \prevdepth0pt\fi
\@authoraddress
\@date
\begin{list}{}{\leftmargin=0.10753\textwidth \rightmargin=\leftmargin
\itemsep=1pc\partopsep=-1pc}
\item\@abstract
\item\@pacs
\end{list}
}

\catcode`\@=12
\begin{document}
\title{Vacuum-driven Metamorphosis}
\author{Leonard Parker\thanks{Electronic address:
leonard@uwm.edu} and Alpan Raval\thanks{Electronic address: raval@uwm.edu}\\
{\small {\it Department of Physics, University of Wisconsin-Milwaukee, P.O. 
Box 413, Milwaukee, WI 53201.}}}
\abstract
{\small We show that nonperturbative vacuum effects can produce
a vacuum-driven transition from a
matter-dominated universe to one in which the effective equation of
state is that of {\it radiation} plus cosmological constant. The actual
material content of the universe after the transition remains that of
non-relativistic matter.
This metamorphosis of the equation of state can be traced to
nonperturbative vacuum effects that cause the scalar curvature to
remain nearly constant at a well-defined value after the transition,
and is responsible for the observed acceleration of the recent
expansion of the universe.}

\maketitle2

\section{Introduction}

Recent observations of high-redshift Type Ia supernovae (SNe-Ia)
\cite{perl,reiss} indicate that the expansion of the universe is 
accelerating at the
present time. Attempts to explain this effect include a revival of the
cosmological constant term in Einstein's equations \cite{sahni,bahcall} as
well as the hypothesized existence of a classical scalar field with a
potential term (quintessence \cite{wang}). In previous work
\cite{parrav1,parrav2}, we show that a free quantized massive
scalar field can account for the SNe-Ia observations in a new
way. Nonperturbative vacuum contributions to the effective action
of such a field \cite{partom} lead to a cosmological solution in which 
the scalar
curvature has a constant value after a cosmic time $t_j$ that depends
on the mass of the scalar field and its curvature coupling. This
solution implies an accelerating universe at the present time and
gives a good fit
to SNe-Ia data for a particle mass parameter of about $10^{-33}$
eV, without the necessity of introducing a cosmological constant term into
Einstein's equations. Such a low mass parameter may arise from
pseudo Nambu-Goldstone bosons \cite{bosons}. A graviton of such a small 
non-zero mass would be expected to produce similar effects.

In Ref. \cite{parrav2}, we also show that our cosmological model gives
light-element abundances in agreement with standard big-bang
nucleosynthesis and that it is consistent
with present data on the small angular scale fluctuations  of the cosmic
microwave background radiation (CMBR), which tend to favor spatial
flatness. 
Our model does not suffer from a fine-tuning problem because, for the
allowed range of the mass parameter, the probability is high that the
matter and vacuum energy densities are of the same order of magnitude
at the present time.

The main point of this letter is to elucidate a remarkable feature of
the equation of state of our model, namely, that the equation of state
undergoes a {\it metamorphosis} from an equation of state dominated by
pressureless matter (without a cosmological constant) to an effective 
equation of state that can be described in classical terms as that of
mixed radiation and cosmological constant. As we show, this metamorphosis is
initiated by the quantum vacuum terms which grow rapidly after time
$t_j$ and combine with pressureless matter to give an effective stress
tensor identical to that of radiation plus cosmological constant.

In the next section we summarize the main features of our model, and
in Section III we analyze the equation of state. Our conclusions are
given in Section IV.
       
\section{Summary of our model}

We consider a free, massive quantized scalar field 
of inverse Compton wavelength $m$, 
and curvature coupling $\xi$. 
The effective action for gravity coupled to such a field is obtained by
integrating out the vacuum fluctuations of 
the field \cite{parrav1,partom}.
This effective action is 
the simplest one that gives
the standard trace anomaly in the massless-conformally-coupled limit, and
contains the nonperturbative sum (in arbitrary dimensions) 
of {\it all} terms in the 
propagator having at
least one factor of the  scalar curvature, $R$. 
The trace of the Einstein equations, obtained by variation of
this effective action with respect to the metric tensor takes the
following form in a
Friedmann-Robertson-Walker (FRW) spacetime (in units such that $c=1$),
with {\it zero} cosmological constant\cite{parrav1,parrav2}:
\bea
\label{one}
R + \frac{T_{cl}}{2\k_o} &=& \frac{\hbar m^2}{32 \pi^2
\k_o}\left\{\vphantom{\frac{m^2}{M^2}} (m^2 +\xib R)\ln 
\mid 1+\xib R m^{-2}\mid \right.\nn \\
& &-\left. \frac{m^2\xib
R}{m^2 +\xib R}\left(1+{3\over 2}\xib \frac{R}{m^{2}}\right.\right. \nn \\
& &+ \left.\left.\ha \xib^2 
\frac{R^2}{m^{4}} (\xib^2 - (1080)^{-1}) +v \right)\right\}, 
\tea
where $T_{cl}$ is the trace of the stress tensor of  classical, 
perfect fluid matter, 
$\k_o \equiv (16\pi G)^{-1}$
($G$ is Newton's constant), $\xib \equiv \xi -1/6$, and $v \equiv 
(R^2/4 - R_{\m \n}R^{\m \n})/(180 m^4)$ is a curvature invariant
that vanishes in de Sitter space.

As noted earlier, $m$ is the inverse Compton wavelength of the
field. It is related to the actual mass of the field by $m_{\rm actual} =
\hbar m$. Equation (\ref{one}) above is nonperturbative in $R$ because it
contains terms that involve an infinite sum of powers of $R$. However, for a
sufficiently low mass, it is possible to treat 
$m_{\rm actual}^2/m_{Pl}^2 \equiv \hbar m^2/(16 \pi \k_o)$ (where $m_{Pl}$ is
the Planck mass) as a small
parameter and expand perturbatively in this
parameter, as we do in obtaining the solution described below.   

In Refs. \cite{parrav1} and \cite{parrav2}, we show that, for a sufficiently
low mass,
in an expanding FRW universe 
the quantum contributions to the Einstein equations become significant
at a time $t_j$, when the density of classical matter, $\rho_m$, 
has decreased to
a value given by
\be
\label{rhcond}
\rho_m(t_j) = 2\k_0 \mb^2,  
\te
where
\be
\mb^2 \equiv m^2/(-\xib).
\te
We find that the time $t_j$ occurs in the
matter-dominated stage of the evolution.
Furthermore for $t > t_j$ we find that the scalar curvature, $R$, 
remains constant to excellent approximation at the
value $\mb^2$. For $t<t_j$, the
quantum contributions to the Einstein equations are negligible and the
scale factor is that of a
matter-dominated FRW universe.
Then, Eq. (\ref{rhcond}) implies that, in a spatially flat universe
with line element $ds^2 = -dt^2 + a(t)^2(dx^2+dy^2+dz^2)$, one has
\be
t_j = (2/\sqrt{3})\mb^{-1},~~~~H(t_j) = \mb/\sqrt{3},
\te
where $H(t_j)$ is the Hubble constant at $t_j$.

As shown in \cite{parrav2}, the condition of the constancy of the
scalar curvature after $t_j$ leads to a solution for the scale factor 
that can be
joined, with continuous first and second derivatives (i.e., in a $C^2$
manner), to
the matter-dominated solution for $t < t_j$. The scale factor is then
given by
\bea
\label{scfactor}
a(t) &=& a(t_j) \sqrt{\sinh\left.\left(\frac{t\mb}{\sqrt{3}} - \a
\right) \right/\sinh\left(\frac{2}{3}-\alpha\right)},~~~~t>t_j,\nn\\
&=& a(t_j)\left(\sqrt{3}\,\mb t/2\right)^{2/3},~~~~t<t_j,
\tea
with
\be
\a = 2/3 - \tanh^{-1}(1/2) \simeq 0.117.
\te
We find that the above solution also satisfies, up to terms of order
$m_{\rm actual}/m_{Pl}$, the one remaining
independent Einstein equation in a FRW universe, which can be taken to
be the time-time component $G_{00} = (2\k_o)^{-1}T_{00}$, where 
$G_{\m \n}$ is the Einstein tensor. This
equation takes the form, with zero cosmological constant,
\begin{figure}[htb]
\centering
\leavevmode
\epsfysize=5.in\epsffile{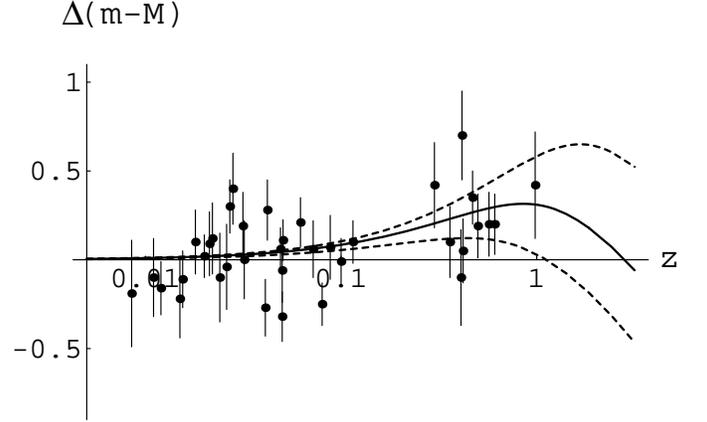}
\caption{A plot of the difference between apparent and absolute
magnitudes, 
as functions of redshift $z$, normalized to an open universe with
$\O_0 =0.2$ and zero cosmological constant. The points with vertical
error bars represent SNe-Ia data obtained from
Ref. [2]. 
The two
dashed curves represent the values (a) $\mb/h = 6.40 \times
10^{-33}$ eV (lower dashed curve), and (b) $\mb/h
= 7.25 \times 10^{-33}$ eV (upper dashed curve). 
The solid curve 
represents the intermediate value (c) $\mb/h = 6.93 \times 10^{-33}$ eV.}
\label{figu1}
\end{figure}

\bea
\label{zzcomp}
k_o G_{00} &=& \ha \rho_m - \frac{\hbar}{64\pi^2}\left\{ \frac{\xib
R_{00}}{m^2 +\xib R}\left(m^4 + 2 m^2\xib R \right.\right.\nn \\
& &+ \left.\left.\frac{R_{\a \b}R^{\a
\b}}{90} +R^2 \left(\xib^2 -{1\over 270}\right)\right) - 3 \xib^2
R\,R_{00}\right.\nn \\
& & +\left.\vphantom{\frac{\xib}{R}} m^2 \xib
G_{00}\right\} 
- \frac{\hbar}{64\pi^2}\ln \left|1 + \frac{\xib R}{m^2}\right|
\left\{- \frac{m^4
g_{00}}{2}\right.\nn \\
& &+ 2 m^2 \xib G_{00} -\frac{g_{00}R^2}{2}\left(\xib^2 +
\frac{1}{90}\right) \nn \\
& &+ \left.\frac{1}{90}g_{00}R_{\a \b}R^{\a \b} + 2\xib^2
R\,R_{00} - \frac{2}{45}{R_0}^{\a}R_{0\a}\right\}.
\tea
To verify that Eq. (\ref{scfactor}) is indeed a solution of the
above equation for $t>t_j$, we note that, when $R$ is very close to
the value $\mb^2$, the dominant terms in the right hand side of 
Eq. (\ref{zzcomp}) are those that have a factor of $m^2 + \xib R$ in
the denominator. Keeping these terms and substituting for the various
curvature quantities derived from Eq. (\ref{scfactor}), it is
straightforward to check that Eq. (\ref{scfactor}) satisfies
Eq. (\ref{zzcomp}) up to terms of order $m_{\rm actual}/m_{Pl}$. 

The solution in Eq. (\ref{scfactor}) corresponds to a universe that 
is accelerating
(i.e., has negative deceleration parameter) for $t >
\sqrt{3}\mb^{-1}(\a + \tanh^{-1}(2^{-1/2})) \simeq 1.50\, t_j$. 
This solution
gives a good fit to the SNe-Ia data, for the mass range
\be
\label{mrange}
6.40 \times 10^{-33} \,{\rm eV} < \left(\frac{\mb}{h}\right) < 7.25 
\times 10^{-33}\, {\rm eV},
\te
where
$h$ is the present value of the Hubble constant, measured as a
dimensionless fraction of the value $100$ km/(s Mpc). In Fig. \ref{figu1}, we 
give plots from \cite{parrav2} of the difference between apparent and 
absolute magnitudes
of sources as functions of the redshift $z$. The two dashed curves
represent the two extremal values of the range (\ref{mrange}). The
solid curve represents an intermediate value of $m_h$, and the data
points represent SNe-Ia data acquired from Ref. \cite{reiss}.

In \cite{parrav2}, we also show that the ratio of  matter density
to critical density at the present time, $\O_0$, is a function of
the single parameter $\mb/h$ and turns out to have the range $0.58 >
\O_0 > 0.15$ for the range of values of Eq. (\ref{mrange}). For
the same range of values, the age of the universe $t_0$ lies in
the range $8.10\,h^{-1}$ Gyr $<t_0<$ $12.2\,h^{-1}$ Gyr.  

\section{Equation of State} 

Although our model incorporates the effective action of a quantum field,
it admits a simple, classical representation.  Indeed, the scale
factor (\ref{scfactor}) may be used to find the total effective energy
density $\rho$ and pressure $p$ of vacuum plus matter by directly 
computing the Einstein tensor. 
We obtain, for $t>t_j$,
\bea
\label{rh}
\rho(t) &=& 2\k_0 G_{00} = (\k_0 \mb^2/2) \coth^2\left(t \mb/\sqrt{3}
- \a\right)\nn \\
& &= (3/2) \k_o \mb^2 \left(a(t)/a(t_j)\right)^{-4} +
(1/2)\k_o \mb^2\\
\label{p}
p(t) &=& 2\k_0 a(t)^{-2}G_{ii} = (\k_0 \mb^2/6) \coth^2\left(t \mb/\sqrt{3}
- \a\right) \nn \\
& &- (2/3)\k_0 \mb^2.
\tea
The effective equation of state for $t>t_j$ is therefore
\be
\label{eost}
p = (1/3)\rho - (2/3)\k_0 \mb^2,
\te
which is identical to the equation of state for a  classical model
consisting of radiation plus 
cosmological constant. 
In our
model the equation of state of pressureless matter and the 
equation of state of quantum vacuum terms combine in a manner so as to
effectively appear as a sum of radiation and cosmological
constant equations of state. Our model differs, even at the
classical level, from the usual mixed matter-cosmological constant
model because (i) for $t<t_j$ the effective cosmological constant
vanishes, 
and (ii) for $t>t_j$ vacuum contributions transmute the effective
equation of state into that of {\it radiation} (rather than
pressureless matter) plus cosmological constant; this surprising
metamorphosis is a result of the near-constancy of the scalar
curvature, which causes certain terms in $T_{\mu \nu}$ to take the
form of an effective cosmological constant term in Einstein's
equations. In a {\it general} spacetime, these terms do {\it not} have
the form of a cosmological constant term.

The equation of state for the quantum vacuum terms {\it alone} may be
inferred from Eqs. (\ref{rh}) and (\ref{p}), and from the fact that
the density of pressureless matter is given by
\bea
\rho_m (t) &=& \rho_m(t_j)\,(a(t_j)/a(t))^3  \nn \\
&=& 2\k_0 \mb^2 \left(\frac{\sinh(2/3 -
\a)}{\sinh(t\mb/\sqrt{3}-\a)}\right)^{3/2},
\tea
where Eqs. (\ref{rhcond}) and (\ref{scfactor}) have been used 
to arrive at the second equality. The quantum vacuum energy density
$\rho_V$ and pressure $p_V$ then follow, for $t>t_j$, as
\bea
\rho_V(t) &=& \rho(t) - \rho_m(t) = \frac{\k_0
\mb^2}{2}\left[\coth^2\left(t \mb/\sqrt{3} - \a\right)\right.\nn \\
& & - \left. 4
\left(\frac{\sinh(2/3 -
\a)}{\sinh(t\mb/\sqrt{3}-\a)}\right)^{3/2}\right]\\
p_V(t) &=& p(t),     
\tea
with $p(t)$ given by Eq. (\ref{p}). The above equations
show that for $t>t_j$ the vacuum energy density is positive, while the
vacuum pressure is negative. As stated earlier, the vacuum
terms are negligible for $t<t_j$.

After some straightforward
algebra, we find the equation of state for the quantum vacuum
terms:
\be
\label{eosv}
\rho_V = 3 p_V + 2\mb^2 \k_0\left[1-\left(1+
2p_V/(\k_0\mb^2)\right)^{3/4}\right].
\te
This vacuum equation of state joins continously to the equation of state
$\rho_V = p_V = 0$ at $t=t_j$ and asymptotes to the
pure cosmological constant equation of state $\rho_V =
-p_V$ as $t \rightarrow \infty$. Equation (\ref{eosv}) shows that 
the quantum vacuum stress
tensor in our model is parametrized by the single parameter, $\mb$, and
is different from that of a pure cosmological
constant. 

\section{Conclusions}

In this letter, we have outlined a cosmological model 
that includes nonperturbative quantum vacuum effects of a
renormalizable free field of very low mass in curved
spacetime. 
The only adjustable parameter,
determined by observation, is the mass scale $\mb$ of the proposed
particle. With $\mb$ in the range of Eq. (\ref{mrange}), the model is
in agreement
with the observed light-element abundances, CMBR fluctuation spectrum,
age of the universe, and the magnitude-redshift relation of Type Ia 
supernovae.

In terms of the equation of state, our model admits a
simple description that could also be a generic feature of theories
other than the one considered here. At a time $t_j$, $m^2 +\xib R$
first becomes small and terms in the quantum
stress tensor
become large, causing the scalar curvature
after $t_j$ to take the nearly constant value $\mb^2 = -m^2/\xib$ (for
$\xib<0$). The constancy of $R$ effectively forces the combined
equation of state of
quantum vacuum terms and pressureless matter to undergo a
metamorphosis to that of  radiation plus cosmological
constant. This effect explains the 
observations regarding the acceleration of the recent expansion of the
universe. Other mechanisms that cause the scalar curvature to remain
constant would give rise to a similar metamorphic
behavior of the equation of state at the phenomenological
level. Nonperturbative terms in the vacuum stress tensor seem to
provide the simplest example of such a mechanism.

Finally, we note that the qunatum vacuum terms would also alter
density inhomogeneities that existed at time $t_j$ because these terms
would remain insignificant in regions having average density at $t_j$
large with respect to $\rho_m (t_j) \equiv 2 \mb^2\k_o$. This may
eventually provide a further observational test of our model.  \\
\bigskip\\
\noindent {\bf Acknowledgements}\\

\noindent This work was supported by NSF grant PHY-9507740.

\end{document}